\documentclass[jkps,preprint, preprintnumber,fleqn,showkeys]{revtex4}
\usepackage{graphicx}
\usepackage{amssymb}
\usepackage{amsmath}
\usepackage{bm}

\usepackage{xfp}
\newcommand\SupplementaryMaterials{%
  \xdef\presupfigures{\arabic{figure}}
  \xdef\presupsections{\arabic{section}}
  \xdef\presupsections{\arabic{table}}
  \renewcommand\thefigure{S\fpeval{\arabic{figure}-\presupfigures}}
  \renewcommand\thesection{S\fpeval{\arabic{section}-\presupsections}}
  \renewcommand\thetable{S\fpeval{\arabic{table}-\presupsections}}
}

\usepackage[normalem]{ulem}

\newcommand{\fref}[1] {{Figure \ref{#1}}}
\newcommand{\sref}[1] {{Section \ref{#1}}}

\usepackage{xcolor}

\newcommand{\ybyso}{{$^{171}\mathrm{Yb}^{3+}$:$\mathrm{Y}_2\mathrm{SiO}_5$}}
\newcommand{\yso}{{$\mathrm{Y}_2\mathrm{SiO}_5$}}
\newcommand{\cflow}{{$^2$F$_{7/2}$(0)}}
\newcommand{\cfhigh}{{$^2$F$_{5/2}$(0)}}
\newcommand{\suppl}{{supplementary materials}}
\newcommand{\ybion}{{$^{171}\mathrm{Yb}^{3+}$}}

\newcommand{\reion}{{RE$^{3+}$}}

\newcommand{\sitea}{{site I}}
\newcommand{\siteb}{{site II}}

\begin{document}
\setcounter{page}{1}
\title[]{Optical and Raman spectroscopies of  {\ybyso} hyperfine structure for application toward microwave-to-optical transducer }
\author{Hee-Jin \surname{Lim}}
\email{heejin.lim@kriss.re.kr}
\author{Gahyun \surname{Choi}}
\author{KeeSuk \surname{Hong}}

\affiliation{Korea research institute of standards and science (KRISS), Daejeon
34113, Korea}


\begin{abstract}
This study analyzed the optical techniques for high resolution, low-noise spectroscopy of  a hyperfine structure (HFS) made of ytterbium-isotope 171 ions ({\ybyso}).
Large energy {spacings in} {\ybion} {are} advantageous for {spin-state} preparations of quantum memory and {construction of a} transducer, {thereby promoting} {the simultaneous} stable {control} of {the} optical frequencies of lasers  over a wide range of 3 {GHz}.
We also built our own {2.7-K} cryogenic system for {optical,} radio-wave-assisted spectroscopy.
We attained to high resolution and sensitivity both in pump-probe saturation spectroscopy (PPS) and Raman heterodyne spectroscopy (RHS).
{Our frequency-stabilized PPS achieved} a high-resolution spectrum of {the HFS}, {whereas} {our setup of RHS} enabled {the efficient detection of} paramagnetic spin resonance efficiently for {a} wide range of {radio frequencies}. 
As the underlying Raman process is {an} up-converting transduction, we {present} {the optimization of the} sensitivity of Raman heterodyne detections {by} selecting the best crystal orientation and  efficient radio-wave coupling {in} future applications {toward} photon {transducers}.
\end{abstract}

\preprint{J. Korean Phys. Soc. 84, 50--58}

\maketitle

\section{INTRODUCTION}

Spin ensembles of rare earth ion ({\reion}) \cite{RN341,marino:pastel-00746050}  in optical crystals {have been actively studied as} materials for {application in} quantum memory \cite{RN7,RN408} {with} long coherence storage time \cite{RN96} and {in} photon {transducers} \cite{RN560, RN61}.
{This material} features  coherent interactions both with optical \cite{RN1373} and microwave photons \cite{RN360, RN33}.
Coherence transfers {between} spins \cite{RN769} and between optical {photons} and nuclear spins \cite{RN11}, which are {crucial in the} access protocol {of} quantum memory,
have been demonstrated with the material.

Raman heterodyne {spectroscopy, which} exploits coherence transfer from a spin transition to an optical {transition,} has been {widely used for reading} coherent {spins} \cite{RN96}, enabling {the optical observation} of spin transient oscillations.\cite{RN46, Holliday:90}
Recently, 
{complementary methods of using nano-photonics \cite{RN83, RN89} and microwave resonators in addition to optical resonators \cite{RN91, RN83, RN36} were proposed to enhance the efficiency of such coherent up-conversion process.}
However,  one important {task} is {still} remained for high-efficiency, high-fidelity state preparation.\cite{RN65}
{The preparation of optical methods in the case of spin-ensemble quantum memory, comprising of complete depletion of both target states and a optical frequency domain for address and followed by re-pumping the nuclear spin states back to the optical domain with addressed frequencies, can be limited for fine manipulations in some {\reion}'s of dense spin structures, due to large optical inhomogeneous broadening that commonly exists for {\reion} doped in solid crystals.}\cite{RN43, RN12}

Among {the many known} {\reion}, ytterbium-isotope-171 ion {doped in} yttrium orthosilicate ({\ybyso}) has been expected to {overcome the limitation}.\cite{PhysRevX.10.031060}
{\ybion} has {a} simple hyperfine structure {(HFS)} {owing} to {a} small quantum number of nuclear spin ($I=1/2$), and {also} large energy {spacing} from hundreds {of} megahertz (MHz) to a few gigahertz (GHz).\cite{RN57, RN56}
{Its HFS} is widely ranged over 3 GHz {in optical frequency}, as {reported} in this article.
Despite {such a} wide range, stability and precision of frequency are still required for  addressing {the spin state} in quantum memory applications.
{In the case of} {\ybyso}, {we must determine how to} control {the} optical frequencies of lasers with high precision or high repeatability.  {In addition, a method must be devised to achieve} good stability {with free or dynamic selection in wide ranges.}

{This study presents} a control system for optical frequencies. {The system was} applied to precise spectroscopies for {\ybion} with a laboratory-designed {low-temperature} setup.
Our system {can simultaneously control} two lasers, and maintains their optical frequencies {locked} using high speed feedbacks to a cavity reference and a radio-frequency (RF) source.

We performed two spectroscopies, {namely,} pump-probe saturation spectroscopy and Raman heterodyne spectroscopy.\cite{RN46}
These two spectroscopies were done in 2.7 kelvins (K), where long coherence time ($T_2$) is allowed for various quantum memory experiments.\cite{RN16, RN8}
{Using pump-probe saturation spectroscopy,} we achieved high-resolution and high-sensitivity {spectra} of the {\ybion} {HFS}.
{The} results of {the} hyperfine resolution are discussed in \sref{sec:HFS}.

Using {a} cryostat designed both for Raman heterodyne spectroscopy and vibration isolations, we succeeded {in directly detecting} spin resonance {using} the optical method.
We optimized {the intensity of the} Raman heterodyne {signals} {using} a crystal orientation to maximize {the} coupling strength between {radio-waves} and spin {transitions} {based on} efficient feed-line circuits for {the} radio-waves.
{We devised} an impedance-matched coplanar waveguide on {a} copper printed-circuit-board, {which could be easily} replaced  with {a} superconducting resonator \cite{RN36} {in} future works {to} challenge {the} high-efficiency of photon transduction.
{The} results of {the} instrumental optimization {of sensitivity are} discussed in \sref{sec:RHS}.

\section{Spectroscopy apparatus and methods}
\subsection{Low-temperature setup for Raman heterodyne spectroscopy}

\begin{figure}[h]
\includegraphics[width=12.0cm]{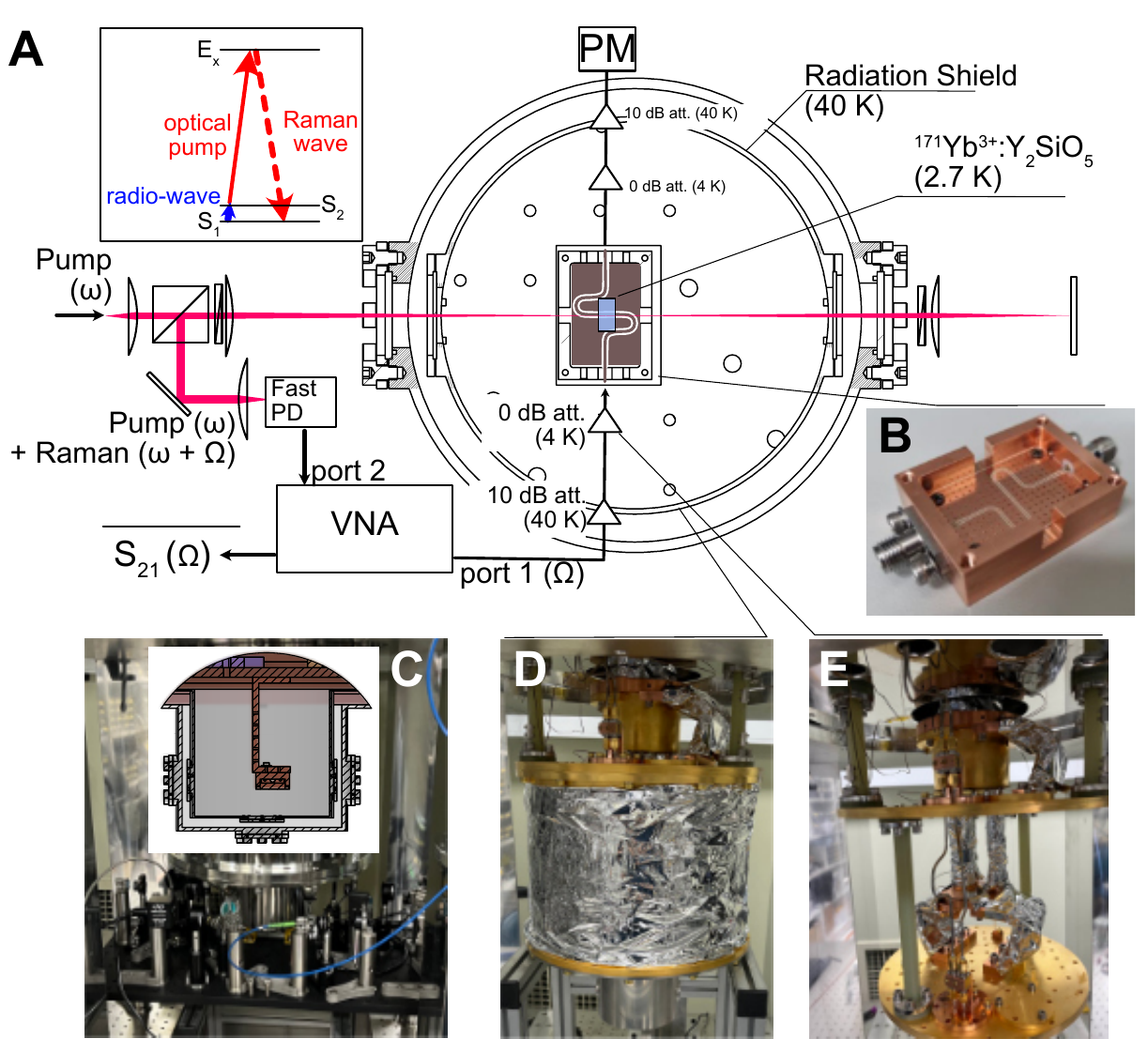}
\caption{(Color online) Experimental setup for low-temperature Raman heterodyne measurements.
{\bf (a)} Schematic  of the setup for optical- and radio-frequency interactions of {\ybyso}  in refrigeration.
{\bf (b)} Microwave coplanar waveguide in a copper box for feeding a radio-waves in {\ybyso}.
{\bf (c,d)} Thermal-radiation-shield apparatus for optical experiments.
{\bf (e)} Thermally protected radio-frequency feed lines.
} \label{fig:RHS}
\end{figure}

 Raman heterodyne spectroscopy is an optical method {used to} detect {the}  spin resonance {of radio-waves}.\cite{RN46, Holliday:90}
This method requires {the collection of} heterodyne beats  between the optical pump and the Raman {waves}.
{In particular,} the generation of Raman {waves} {can be} enhanced by the spin resonance of {radio waves} (transition {of} $S_1 \rightarrow S_2$) within a cyclic process completed by {an} optical pump ($S_2 \rightarrow E_x$) and an emission of Raman {waves} ($E_x \rightarrow S_1$), {as shown in the} inset {of} {\fref{fig:RHS}(a)}.
In this process, the enhanced heterodyne beat should have the same frequency {as that of} the spin transition.
Our {simple} setup for the heterodyne detection is {depicted in} {\fref{fig:RHS}(a)}, {where} detection can be {minimally} {performed using} a vector network analyzer (VNA) and a fast photodiode of 10-GHz bandwidth.
VNA is a useful tool {for generating} and {receiving} radio {waves}, producing a spectrum of the transmission {S-parameter} ($\mathrm{S}_{21}$) of the generated frequency of radio {waves} ($\Omega$).

However, refrigeration to below 4 K is required to ease {the} rate of spin relaxation ($T_1^{-1}$) toward {the} thermal equilibrium population of $S_1$ and $S_2$ .
The main {source} of relaxation in spin ensembles of rare ion ({\reion}) in {\yso} {are the} phonon interactions \cite{RN341}, {which display a} more {intensive impact through} two phonon processes at higher temperatures.\cite{RN761}
{Studies have shown} that the spin relaxation rate of {\ybyso} can be lower than 10 Hz at $< 4$ K.\cite{RN52}
{Therefore,} our setup {was devised to refrigerate}  {\yso} to $< 2.7$ K {by} using a pulse tube cryocooler (Cryomech PT405).
\fref{fig:RHS}(c-e) shows the internal structures of {our designed} cryostat.
To isolate thermal radiation through optical windows, we {installed} cold windows at {the} 30-K radiation shield.
{In addition, we used} semi-rigid coaxial cables made of stainless steel to deliver a high-frequency radio waves to {\ybyso}.
{Futhermore,} low-stiffness copper braids wound with multilayer insulation films were used to isolate vibrations while allowing a large cooling capacity.

For an efficient RF coupling with {respect to} {\ybion}, we {installed} a cube of {\ybyso} crystal on a printed circuit board, {in which} a microwave coplanar waveguide {with a matched} characteristic impedance {was} inscribed. 
{This radio-wave setup is}  protected from electromagnetic noises by {using} a copper box, {as shown in} \fref{fig:RHS}(b).

{As} the RF magnetic field is formed locally around the {central} conductor of {the} coplanar waveguide,\cite{RN917} {the} Raman generations enhanced by resonant radio {waves} can {occur} for {the} {\ybion} that locates {near this} conductor.
{Therefore,} {the optimizations involves an} increase {in the} optical power density {in this} region {until} photocurrents and electrical amplification in the fast photodiode do not saturate.
In our setup, the pump laser beam of 1-mW power {was} focused {toward} an area of $\sim$ 0.1-mm diameter with slow beam divergence, located near the {central} conductor, and adjusted to avoid surface scattering.
We {ensured that} the beam direction was parallel to the waveguide direction to {achieve} maximum overlap between the optical path and the RF magnetic field.
The {\yso} crystal was {carefully oriented} on the waveguide {such that} the crystallographic $b$-axis {was} aligned perpendicular to the waveguide direction.
This crystal orientation {allows the maximization of the} coupling strength between the radio waves and spin transitions, benefited by anisotropy of magnetic susceptibility.
{The theoretical discussion is detailed} in \sref{sec:RHS}.

{For} our sample, {we ordered to substitute} 50 ppm of crystallographic yttrium sites ({\sitea} and {\siteb}) {with} isotopically purified {\ybion}.
Considering that {the} two crystallographic sites in {\yso} have different optical transition frequencies and HFSs, 
the population in {the} interaction should be counted only for a specific site.
The total population excited by the pump laser is {approximately} 3.8 $\times 10^{13}$ for {the} length of {the} optical path inside {\yso} {used in this study} (10 mm).

\subsection{Optical frequency control system}

\begin{figure}[h]
\includegraphics[width=12.0cm]{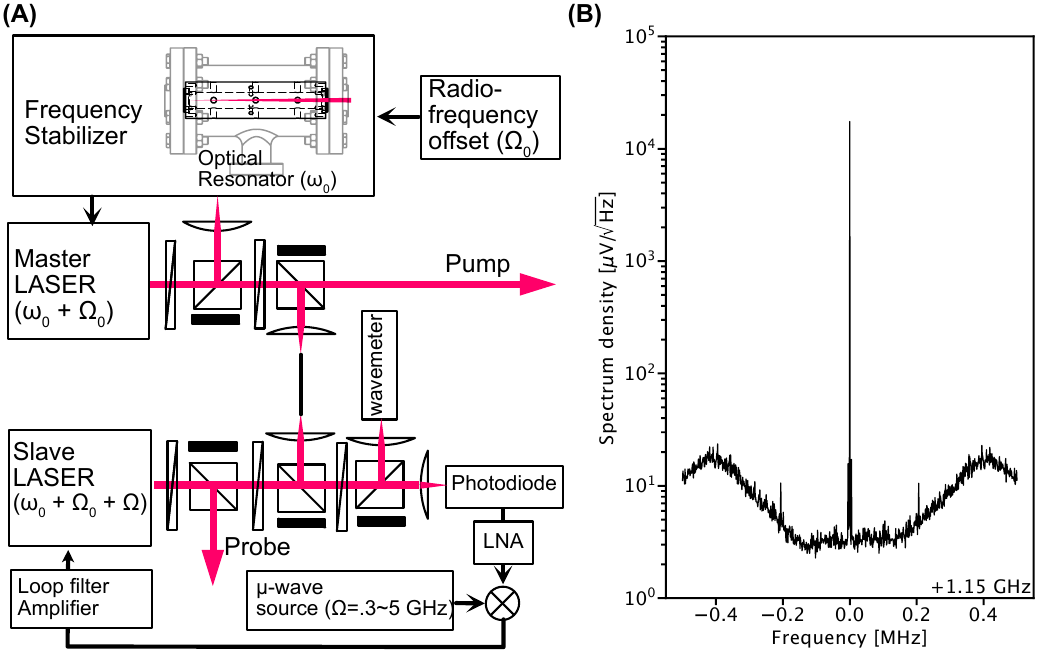}
\caption{
{\bf (a)} Laser-frequency stabilizing system.
{\bf (b)} Beat spectrum of interference signals between the master and the slave lasers.
The horizontal axis {represents} the beat frequency relative to 1.15 GHz, {which} corresponds to the difference {in the} optical frequencies of {the} lasers.
{The spectrum for the main 1.15-GHz signal is detailed as a phase noise chart in the {{\suppl}}.}
}\label{fig:OPLL}
\end{figure}

{The} stabilization of a relative optical frequency {between the} pump (master) and probe (slave) lasers enables {highly} sensitive and repeatable detections, {thus allowing} high precision {when} resolving the {HFS} of {\ybyso}.
{In addition}, {the} stabilization of the master laser {and the} relative frequency of the slave laser allows spin-state addressing with absolute optical frequencies for future quantum memory applications.
Our dual schemes of stabilization also {provide protection} from environmental acoustic noises {at} higher levels.

Our system for optical frequency control for pump and probe lasers combines
{the} optical cavity lock \cite{Thorpe:08} and optical phase lock loop techniques(OPLL) \cite{Fang2017}.
{For an optical cavity reference,} we adopted the electronic side-band {(ESB)} modulation technique \cite{Thorpe:08}, {as depicted by the} the frequency stabilizer box in \fref{fig:OPLL}(a).
This method has a lock point offset by a radio-frequency ($\Omega_0$) from the cavity resonance ($\omega_0$).
Therefore, we could maintain {the stability of} the master laser at {the} desired frequency.
Details of our ESB method are explained in {the} {\suppl}.
{In this study,} we {constructed an} ultra-high-vacuum body for {an} optical cavity with high reflectivity ($>$ 99.9 \% concave mirrors) and achieved a low error density $<$ 10 Hz/Hz$^{1/2}$ {for the locked optical frequency} within {the} measurement bandwidth {of 10 kHz}, {using} a commercial fast PID controller (Toptica FALC pro).

{The} phase detection of the slave laser in our OPLL setup {was} processed {through} generating optical beats and comparing {their} phase with a radio-frequency ($\Omega$) reference, as shown in \fref{fig:OPLL}(a).
{The} electronic signals for the optical beat were generated from an impedance-matched photodiode of 10-GHz bandwidth and were amplified {using} a low-noise amplifier (LNA).
The down-conversion of {the} beat frequency with a mixer driven by the local oscillator (LO) frequency ($\Omega$) plays {the role of} a phase detector, and the converted signal modulates {the} electric currents for the slave laser {through the} filter amplifiers of the PID controller.
We employed {external cavity diode lasers} (ECDL) for both master and slave lasers,
taking advantages of {the} high-speed frequency response ($<$ 200 MHz) to electric currents.
Stable phase conditions in the feedback loop {were observed} for a beat frequency identical to {the} LO {frequency}, {to allow the tuning of the} lock point of relative frequency by adjusting {the} LO frequency.
{Therefore,} the locked frequency of the slave laser was determined by $\Omega$.
{In addition,} we achieved a wide tuning range from 300 MHz to 3 GHz, and a continuous sweep range of 100 MHz.
The second range was limited by a radio-wave synthesizer (because of sweep discontinuity of fractional PLL and internal switches).

The spectrum of optical beats locked {at} 1.15 GHz (\fref{fig:OPLL}(b)) {showed a} high level of noise suppression and frequency accuracy.
Although this feedback loop was designed {to consider} low-phase noise, 
we {optimally} tuned {it} for frequency sweep at the {expense} of phase noise reduction, {still confirming} a frequency error far smaller than 1 kHz.
Note that all frequency synthesizers and receivers, including spectrum analyzer, {were} synchronized {according to the} Rubidium frequency standard (SRS FS725).

\section{Hyperfine structure}\label{sec:HFS}

\begin{figure}[ht!]
\includegraphics[width=12.0cm,angle=180]{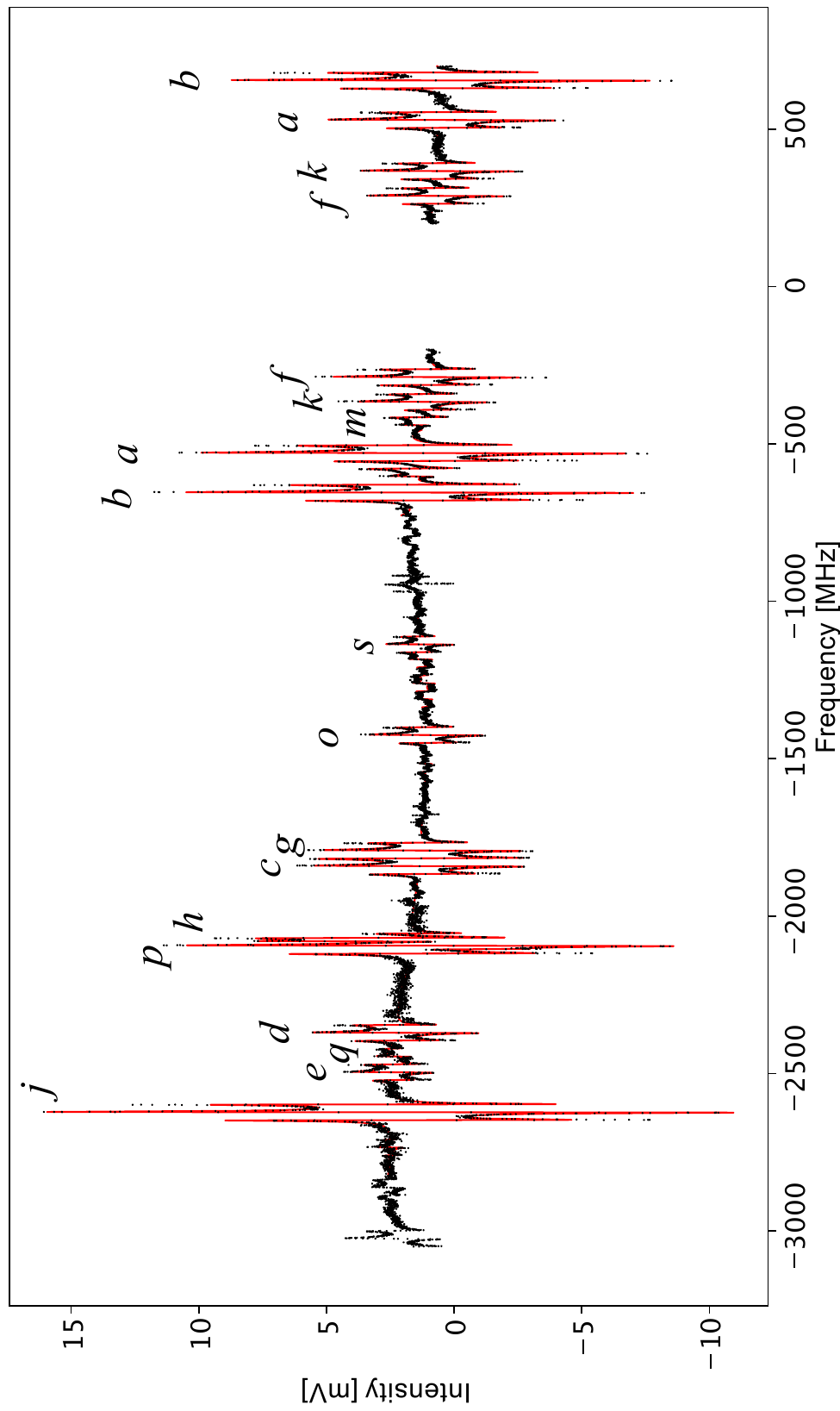}
\caption{Pump-probe saturation spectrum of {\ybyso} (crystallographic {\siteb}) at 2.7 K.
The horizontal axis {represents} the optical frequency difference {between the} pump and probe lasers.
Resonances are {represented} by small letters ({\it a}--{\it s}).
}\label{fig:OH}
\end{figure}

\begin{figure}[h]
\includegraphics[width=9cm]{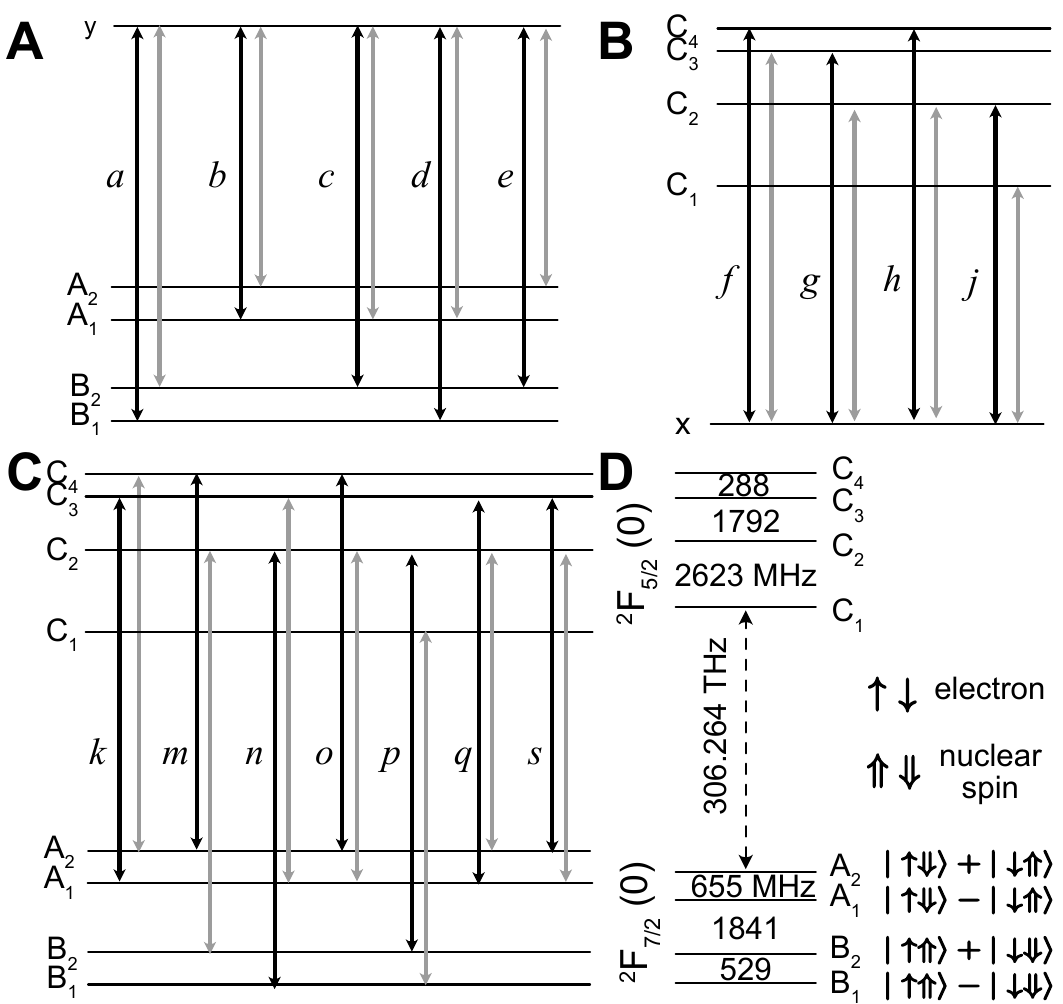}
\caption{Hyperfine structure of {\ybyso} as the origin of pump-probe resonance.
{\bf (a-c)} The origin of resonance signals appearing in \fref{fig:OH}. Black (gray) arrows indicate transitions interacting with the pump (probe) laser.
{\bf (d)} Hyperfine structure of the lowest-lying crystal field (CF) state {\cflow} and the optically accessible {\cfhigh} for {\ybion} at crystallographic yttrium {\siteb} of {\yso}.
}\label{fig:HFS}
\end{figure}

{\ybion} {demonstrates} nuclear spin {with a} quantum number {of} $I = 1/2$,  thus {displaying} a  HFS {in} both  the lowest-lying crystal-field {(CF)} state {\cflow} and in the optically accessible {CF} state {\cfhigh} as shown {by} the energy diagram in \fref{fig:HFS}(d).
The strong coupling strength between {the} electron and nuclear spins {causes} large energy splittings in {the} HFS of {\cflow}.
{The} eigen-states of {\cflow} {are denoted as} $A_{1,2}$ and $B_{1,2}$  and {those of {\cfhigh} are denoted by} $C_{1,2,3,4}$.
The representation of {the} eigen-states with electron ($\uparrow, \downarrow$) and nuclear spins ($\Uparrow, \Downarrow$), as shown in \fref{fig:HFS}(d), depends on {the} setting of {the} quantization axis, and can  slightly {differ} from {those presented in} earlier {studies} \cite{RN75, RN56}.

{Using} our frequency control system {with a} long tuning range and excellent stabilization performance,
we could measure a highly accurate HFS spectrum at 2.7 K, as shown as {\fref{fig:OH}}.
To increase sensitivity, we {adopted} the frequency modulation (FM) technique using a phase-modulating electro-optic modulator.\cite{LEVENSON198829}
The counter-propagating pump and probe beams were laid on the same path and at the same linear polarization in the plane of  $D_1$ - $D_2$ axes of {\yso}.
Because they are distinguished only by their propagating directions, we devised to identify them using a circulator setup of {comprising a} Faraday rotator and {a} calcite polarizer. 
This setup {could choose a polarization to obtain} the maximum optical absorption.
The {power of the} pump (probe) was 10 (0.1) mW {within} area {of a} 1-mm beam diameter.

{As shown in \fref{fig:OH},} pump-probe spectroscopy {displayed many resonance signals within the spectrum}.
The pump frequency  for {\ybion} at crystal {\siteb} of yttrium in {\yso} was selected as 306.2685 THz, {excited simultaneously} from $\{A_1, A_2\}$ to $\{C_3, C_4\}$ and from $\{B_1, B_2\}$ to $C_2$ to recirculate spin state populations.
The horizontal axis of {the} spectrum {displays} the relative frequency of the probe laser to the pump laser.
The values of {the} half-width-half-maximum linewidth {were} measured between 1 and 1.5 MHz.
The linewidths were attributed to inhomogeneous broadening inside {\cflow} and inside {\cfhigh}, affected by a mechanical strain in {\yso}.\cite{RN52,RN56} 
({Details about the} resonance frequency and linewidth {are provided} in {\suppl}.)

The physical origins of resonances ({\it a}--{\it s}) observed in the spectrum in \fref{fig:OH} can be explained with the energy diagrams shown in \fref{fig:HFS}(a-c).
The transitions resonant with pump and probe lasers of configurations {\it a--e} share the same optically excited state.
Therefore, the spectrum positions of {\it a--e}  are coincident with the energy spacing of the {\cflow} spin eigen-states.
From the spectrum positions of {\it f--j}, we extracted the energy spacings for {\cfhigh}.

However, resonances of {\it k--s} are complicated, because the transitions interacting with the pump and probe lasers have no common state, as shown as \fref{fig:HFS}(c).
They exhibited smaller signal intensities, enabled by additional processes of relaxations intermediate between the pump and probe transitions.
Their spectral positions were matched with algebraic relations of energy spacing, as noted in a supplementary table ({\suppl}).
As observed, the blank spaces denote resonances that were weakly detected.

\section{Experimental results of Raman heterodyne spectroscopy}\label{sec:RHS}

\begin{figure}[h]
\includegraphics[width=15cm]{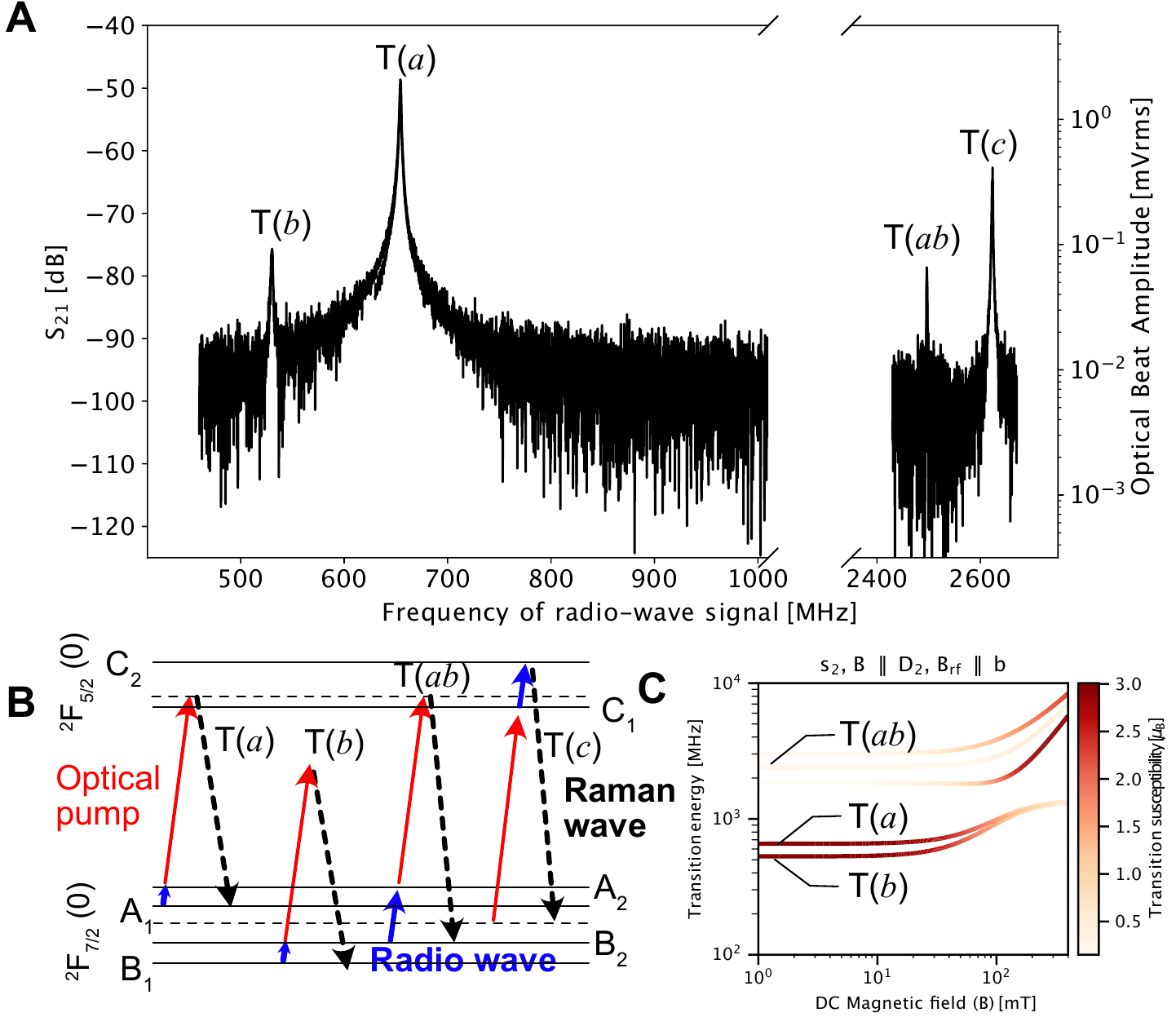}
\caption{Raman heterodyne spectra of {\ybyso}- {\siteb} spin resonance.
{\bf (a)}  Horizontal axis denotes the radio-wave frequency. $S_{21}$ denotes the logarithm ratio of the power of a Raman heterodyne beat to the power of radio waves fed into the copper box. RF power is -12 dBm (63 $\mu$W), and the optical power is 1.3 mW with a  0.1-mm  diameter.
{\bf (b)} Energy diagram of the observed spin resonances of Raman heterodyne process. 
{\bf (c)} Spin energies calculated using the spin Hamiltonian of {\cflow}. The color displays the transition magnetic susceptibility normalized by the Bohr magnetron, implying the coupling strength of the radio-wave.}\label{fig:RHA}
\end{figure}

The main results of Raman heterodyne measurements are displayed by the spectrum in \fref{fig:RHA}(a).
In this spectrum, radio-wave resonance were observed for spin transitions of $A_1 \leftrightarrow A_2$, $B_1 \leftrightarrow B_2$, and $A_2 \leftrightarrow B_2$, denoted as T({\it a}), T({\it b}), and T({\it ab}) respectively as in the energy diagram of \fref{fig:RHA}(b).
Futhermore, we observed that the resonant frequencies and linewidth of the radio-waves were identical to those obtained by the pump-probe saturation spectroscopy.

The signal intensity depends on the magnetic susceptibility of the transition and the optical frequencies of the pump laser.
Here, we set the optical frequency of pump laser close to the transition energy of $A_2 \leftrightarrow C_1$, but slightly offset it by 1.5 GHz.
This offset is similar to a full-width-half-maximum width of optical inhomogeneous broadening, and could maximize for maximizing the T({\it a}) signal intensity by avoiding the depletion of $\{A_1, A_2\}$. 
Then, the T({\it a}) signal intensity includes a contribution from the process comprising of a radio-wave driving $A_2 \leftrightarrow A_1$, optical pump $A_1 \rightarrow C_1$, and Raman wave generation $C_1 \rightarrow A_2$, as well as the configuration $\{A_1 \rightarrow A_2, A_2\rightarrow C_1, C_1 \rightarrow A_1\}$ presented in \fref{fig:RHA}(b).
Although the pump laser frequency became close to the $B_2\rightarrow C_1$ excitation by the offset, the optical pumping seems still selective in the excite $\{A_1, A_2\}$ than $\{B_1, B_2\}$.
From a study of optical absorption spectrum (\suppl), $\{B_1, B_2\}$ prefers optical excitations into $\{C_2, C_3, C_4\}$ than into $C_1$, which may depend on our setting of polarizations, or sample characteristics.
In this case, Raman heterodyne detections for $\{B_1, B_2\}$ states requires $> 3$-GHz greater optical frequencies of the pump to be excited to $\{C_2, C_3, C_4\}$.

The weak signal intensity of T({\it c}) is due to a negligibly small magnetic susceptibility of transition, as shown as \fref{fig:RHA}(c).
To theoretically achieve  this value of transition strength, we first calculated the spin-state spacing and its eigen-state from the spin Hamiltonian ($\hat H$) of electron and nuclear spins ($\hat S, \hat I$) using a symmetric hyperfine tensor $A_{ij}$,  symmetric gyromagnetic tensor $g_{ij}$ for $i,j \in \{D_1,D_2,b\}$, Bohr magnetron $\mu_B$, and magnetic field $B$ \cite{PhysRevX.10.031060, RN56, RN57}:
\begin{equation}
\hat H = \mu_B g_{ij} B_i \hat S_j  + A_{ij} \hat S_i \hat I_j.
\end{equation}
With spin eigen-states of $\psi_\alpha \in \{A_1, A_2, B_1, B_2\}$, the transition magnetic susceptibility ($\vec \chi$) can be defined as
\begin{equation}
\vec \chi_{\alpha\beta} = \mu_B \left \langle \psi_\alpha \left | \overleftrightarrow g \cdot \hat{\vec S} - \frac{\vec B_\mathrm{DC} (\vec B_\mathrm{DC} \cdot \overleftrightarrow g \cdot \hat{\vec S})}{|\vec B_\mathrm{DC}|^2} \right | \psi_{\beta}\right \rangle.
\end{equation}
Here, subtraction can correct the transition strength, keeping only the oscillation component of the magnetic dipole induced by a radio-wave magnetic field ($\vec B_\mathrm{rf}$), excluding the DC magnetization component generated by strong magnetic fields ($\vec B_\mathrm{DC}$).\cite{RN52}
The $g_{ij}$ of {\ybyso} displays large anisotropy, such that it increases to $\sim 6$  for a $\vec B$ direction close to $b$-axis of {\yso}, but it can be diminished to $< 1$ for the other directions.\cite{RN57}

By taking advantages of the $g_{ij}$ anisotropy to obtain a large interaction strength between radio-waves and resonant spin transitions, we aligned the $b$-axis parallel to $\vec B_\mathrm{rf}$ of the coplanar waveguide mode.
We expected a high value of $\chi \sim 3 \mu_B$ for $A_1 \leftrightarrow A_2$ and $B_1 \leftrightarrow B_2$ even though they exist in the electron-nuclear mixture under low $B_\mathrm{DC} < 20$ mT, as shown as \fref{fig:RHA}(c).
In this calculation, $\chi$ of the transition $A_2 \leftrightarrow B_2$ of T({\it ab}) was a small value.

Because of the efficient radio-wave coupling effect and optical beat detection, the radio-wave power set for obtaining the spectrum (\fref{fig:RHA}) was -12 dBm (63 $\mu$W) corresponding to 0.08 mT of peak RF magnetic field strength.
Note that this was achieved without the use of an RF amplifier.
The absolute optical power of the generated Raman for T({\it a}) was 2 nW, which was converted from the optical beat amplitude (2 mV$_\mathrm{rms}$) using the RF conversion gain of the photodiode (1200 V/W) and the power of optical LO (1.5 mW).
As we did not use an RF resonator, most of the power of the  radio-waves passed though the waveguide,
and the probability of resonant scattering and absorption of radio-wave photons are very low in this configuration.
Therefore, applying a definition of the conversion efficiency as a Raman photon rate over the driving RF photon flux rate seems early for consideration.
Nevertheless, this paper presents the outcome of crude estimation as 7 $\times 10^{-11}$ (= 2 nW /  306 THz $\times$ 655 MHz / 63 $\mu$W).

Here, T({\it c}) is unexpected because it originates in the transition $C_1 \leftrightarrow C_2$.
Note that we did not theoretically study  $\chi$ for the transition inside {\cfhigh} because of lack of information.
That is also because decisive factors, such as $g_{ij}$ and $A_{ij}$, were studied mainly for the lowest-lying CF state ({\cflow}) by using electron-paramagnetic resonance measurements.
Herein, we report  that transition $C_1 \leftrightarrow C_2$ displays high $\chi$, and spin states in {\cfhigh} can be termed as valuable subjects of optically accessible spin ensembles.

\section{CONCLUSIONS}

The high-resolution and low-noise spectra derived using the proposed system  represent reliable noise rejection performance througth feedback controls for generating optical and radio-wave frequencies.
Using the proposed system, the HFS and spin resonances of {\ybyso} could be measured accurately.
Techniques that we employed to obtain the sensitive spectrum of HFS are suitable for the frequency reference  in the precise quantum memory experiment  with {\ybion} in near future.
The spectroscopy methods developed in this study could facilitate in the control, measurement, and preparation of {\ybion} spin states required by quantum memory and transducer applications.

\begin{figure}[h]
\includegraphics[width=8cm]{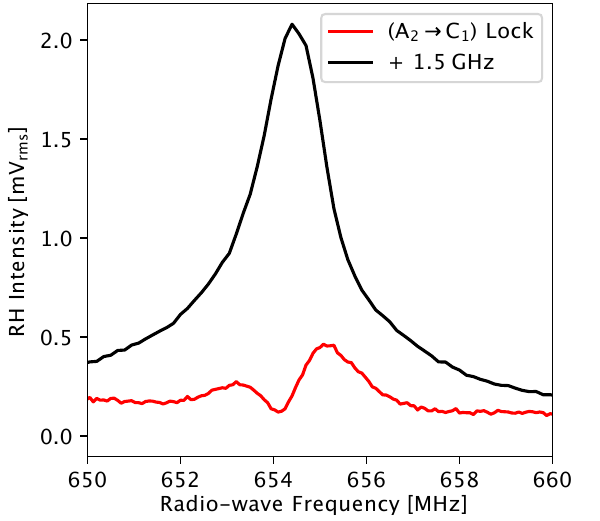}
\caption{Effects of optical pump frequency and its stability  on Raman heterodyne signal. The red curve was obtained with the resonant optical frequency under lock condition. The black curve represents the Raman heterodyne spectrum for a 1.5-GHz detuned optical pump.}\label{fig:depletion}
\end{figure}

The importance of selecting the pump frequency related with the depletion phenomenon were observed in this study.
{\fref{fig:depletion} shows the effect of the phenomenon on Raman heterodyne signals through the suppression of the T({\it a}) signal.}
This could be achieved when the optical pump frequency is resonant with $\{A_1, A_2\} \leftrightarrow C_1$ and in a stable lock condition.
This result can be attributed to the optical pump, $\{A_1, A_2\} \rightarrow C_1$, moving populations toward $\{B_1, B_2 \}$.
Our next study will be constructing a strong spin polarization between $A_1$ and $A_2$.
This is challenging because its transition energy is low to resist forces toward thermal equilibrium at 2.7 K, and the spin states rarely avoid spectral mixing in the optical domain owing to inhomogeneous broadening between {\cflow} and {\cfhigh}.
Nevertheless, as we observed the strong state depletion, we expect to find some clues from our radio-wave assisted optical spectroscopy; future protocols of state preparation may be devised using   resonant radio waves assisted by optical transitions optimally addressed by optical pumping.
The addressing method will be processed by a complete depletion of both $\{A_1, A_2\}$ and address domain and followed by a sharp addressing pump from $\{B_1, B_2\}$ through $\{C_3, C_4\}$, whose relative frequencies of pump lasers are +0 GHz (306.264 THz) with a sweep width of 200 MHz for depletions and +7.13 GHz fixed in stabilization for addressing respectively,
resulting in isolated absorption resonances addressed at +57 MHz ($A_1 \rightarrow C_1$) and -69 MHz ($A_2 \rightarrow C_1$).
Furthermore, one can imagine population burning by applying radio waves resonant with $B_1 \leftrightarrow A_2$ and $B_2 \leftrightarrow A_2$ to increase the population of $A_1$ at the address.

The use of DC magnetic fields to tune spin energy into cavity resonances is a viable option for increasing the interaction strength between spin ensemble and radio waves. 
However, this could further complicate the transducer application, considering the importance of isolation of field and background noises for quantum computing modules.
Even if allowed, there exists strong limitations of field strength and orientation  for low-loss superconducting  resonators \cite{CHSONG}.
Future studies must consider the mechanical or electrical wide-tuning of resonators to achieve optimal transducer application.

\begin{acknowledgments}
This study was supported by the National Research Foundation of South Korea (NRF-2021M3E4A1038018, {\em Transducers between microwave and near-infrared photons based on coherent spin memory}).
\end{acknowledgments}


\newpage
\SupplementaryMaterials
\section*{Supplementary Materials}
\section{Details of the pump-probe spectroscopy results}

\begin{table*}[h!]
\caption{Resonance frequencies shown in the pump-probe saturation spectrum of Figure 3}
\begin{ruledtabular}
\begin{tabular}{c|ccc|ccc}
Index & Frequency & Linewidth & Amplitude & Notes & Pump &Probe \\
& (MHz)& (MHz) & (a.u.) & &  & \\
\hline
{\it f} & 287.76 $\pm$ 0.07 & 0.96 $\pm$ 0.08 & 0.36 $\pm$ 0.01 & C$_3 \rightarrow$ C$_4$   &  x $\leftrightarrow$ C$_3$  & x $\leftrightarrow$ C$_4$ \\ 
{\it f} & -287.80 $\pm$ 0.08 & 1.11 $\pm$ 0.11 &  0.45 $\pm$ 0.01  & C$_4 \rightarrow$ C$_3$   &    x $\leftrightarrow$ C$_4$ & x $\leftrightarrow$ C$_3$ \\
{\it k} &-366.58 $\pm$ 0.10 &1.39 $\pm$ 0.13 & 0.42 $\pm$ 0.01 & f(C$_3 \rightarrow$ C$_4$)-f(A$_1 \rightarrow$ A$_2$)  & A$_1 \leftrightarrow$ C$_3$ &  A$_2\leftrightarrow$ C$_4$ \\
{\it k} &
366.72 $\pm$ 0.08 &
1.07 $\pm$ 0.09 &
0.40 $\pm$ 0.01 &
f(C$_4 \rightarrow$ C$_3$)-f(A$_2\rightarrow$ A$_1$) &
A$_2\leftrightarrow$ C$_4$ &
A$_1\leftrightarrow$ C$_3$  \\
{\it m} &
-415.53 $\pm$ 0.17 &
1.15 $\pm$ 0.27 &
0.25 $\pm$ 0.01 &
 f(A$_2\rightarrow$ B$_2$)-f(C$_4\rightarrow$ C$_2$)  &
A$_2\leftrightarrow$ C$_4$ &
B$_2\leftrightarrow$ C$_2$ \\
{\it a} &
528.83 $\pm$ 0.10 &
1.41 $\pm$ 0.10 &
0.56 $\pm$ 0.01 &
B$_1\rightarrow$ B$_2$ &
B$_2\leftrightarrow$ y &
B$_1\leftrightarrow$ y \\
{\it a} &
-529.19 $\pm$ 0.09 &
1.37 $\pm$ 0.11 &
0.76 $\pm$ 0.01 &
B$_2\rightarrow$ B$_1$ &
B$_1\leftrightarrow$ y &
B$_2\leftrightarrow$ y \\
&
-578.18 $\pm$ 0.27 &
1.21 $\pm$ 0.34 &
0.32 $\pm$ 0.02 &
f(C$_2\rightarrow$ C$_3$)-f(B$_1\rightarrow$ A$_1$) &
B$_1\leftrightarrow$ C$_2$ &
A$_1\leftrightarrow$ C$_3$ \\
{\it b} &
-654.57 $\pm$ 0.08 &
1.12 $\pm$ 0.10 &
0.70 $\pm$ 0.01 &
A$_1\rightarrow$ A$_2$ &
A$_1\leftrightarrow$ y &
A$_2\leftrightarrow$ y \\
{\it b} &
654.68 $\pm$ 0.09 &
0.94 $\pm$ 0.09 &
0.62 $\pm$ 0.01 &
A$_2\rightarrow$ A$_1$ &
A$_2\leftrightarrow$ y &
A$_1\leftrightarrow$ y \\
{\it s} &
-1136.80 $\pm$ 0.10 &
1.08 $\pm$ 0.12 &
0.27 $\pm$ 0.01 &
f(C$_3\rightarrow$ C$_2$)-f(A$_2\rightarrow$ A$_1$) &
A$_2\leftrightarrow$ C$_3$ &
A$_1\leftrightarrow$ C$_2$ \\
 &
-1184.31 $\pm$ 0.24 &
1.03 $\pm$ 0.35 &
0.15 $\pm$ 0.01 &
f(A$_2\rightarrow$ A$_1$)+f(B$_2\rightarrow$ B$_1$) &
 &
 \\
 &
-1260.30 $\pm$ 0.28 &
1.04 $\pm$ 0.50 &
0.15 $\pm$ 0.02 &
f(C$_3\rightarrow$ C$_2$)-f(B$_2\rightarrow$ B$_1$) &
B$_2\leftrightarrow$ C$_3$ &
B$_1\leftrightarrow$ C$_2$ \\
 &
-1310.91 $\pm$ 0.43 &
1.02 $\pm$ 0.58 &
0.12 $\pm$ 0.02 &
f(A$_1\rightarrow$ B$_2$)-f(B$_2\rightarrow$ B$_1$) &
 &
 \\
{\it o} &
-1424.87 $\pm$ 0.09 &
1.42 $\pm$ 0.11 &
0.40 $\pm$ 0.01 &
f(C$_4\rightarrow$ C$_2$)-f(A$_2\rightarrow$ A$_1$) &
A$_2\leftrightarrow$ C$_4$ &
A$_1\leftrightarrow$ C$_2$ \\
 &
-1515.90 $\pm$ 0.18 &
1.01 $\pm$ 0.32 &
0.11 $\pm$ 0.01 &
f(C$_3\rightarrow$ C$_2$)-f(C$_4\rightarrow$ C$_3$) &
 &
 \\
{\it g} &
-1791.80 $\pm$ 0.09 &
1.42 $\pm$ 0.11 &
0.52 $\pm$ 0.01 &
C$_3\rightarrow$ C$_2$ &
x $\leftrightarrow$ C$_3$ &
x $\leftrightarrow$ C$_2$ \\
{\it c} &
-1841.27 $\pm$ 0.07 &
1.43 $\pm$ 0.10 &
0.54 $\pm$ 0.01 &
A$_1\rightarrow$ B$_2$ &
B$_2\leftrightarrow$ y &
A$_1\leftrightarrow$ y \\
{\it h} &
-2080.14 $\pm$ 0.07 &
1.15 $\pm$ 0.10 &
0.45 $\pm$ 0.02 &
C$_4\rightarrow$ C$_2$ &
x $\leftrightarrow$ C$_4$ &
x $\leftrightarrow$ C$_2$ \\
{\it p} &
-2094.14 $\pm$ 0.09 &
1.29 $\pm$ 0.10 &
0.79 $\pm$ 0.01 &
f(C$_2\rightarrow$ C$_1$)-f(B$_2\rightarrow$ B$_1$) &
B$_2\leftrightarrow$ C$_2$ &
B$_1\leftrightarrow$ C$_1$ \\
{\it d} &
-2370.58 $\pm$ 0.09 &
1.22 $\pm$ 0.10 &
0.45 $\pm$ 0.01 &
A$_1\rightarrow$ B$_1$ &
B$_1\leftrightarrow$ y &
A$_1\leftrightarrow$ y \\
{\it q} &
-2445.74 $\pm$ 0.35 &
1.17 $\pm$ 0.39 &
0.20 $\pm$ 0.01 &
f(C$_3\rightarrow$ C$_2$)-f(A$_1\rightarrow$ A$_2$) &
A$_1\leftrightarrow$ C$_3$ &
A$_2\leftrightarrow$ C$_2$ \\
{\it e} &
-2496.13 $\pm$ 0.11 &
1.56 $\pm$ 0.13 &
0.35 $\pm$ 0.01 &
A$_2\rightarrow$ B$_2$ &
B$_2\leftrightarrow$ y &
A$_2\leftrightarrow$ y \\
{\it j} &
-2623.10 $\pm$ 0.09 &
1.03 $\pm$ 0.09 &
0.84 $\pm$ 0.01 &
C$_2\rightarrow$ C$_1$ &
x $\leftrightarrow$ C$_2$ &
x $\leftrightarrow$ C$_1$ \\
\end{tabular}
\end{ruledtabular}
\label{tab:fulldata}
\end{table*}

We fitted each resonance signal in the pump-probe saturation spectrum (Fig. 3) into a line-shape model of the frequency-demodulated signal of a Lorentzian spectral response:
\begin{equation}
I (\omega) = \mathrm{Re} \left[ F^* (\omega - \omega_0) F(\omega - \omega_0 + \nu) - F(\omega - \omega_0) F^* (\omega - \omega_0 - \nu)  \right],
\end{equation}
where $F(\omega) = (\omega + \gamma \sqrt{-1})^{-1}$ with a linewidth factor $\gamma$, phase-modulation frequency $\nu$, and central frequency $\omega_0$, according to the frequency-modulation spectroscopy \cite{LEVENSON198829}.
The values in the {\it frequency} and {\it linewidth} columns display the parameters of $\omega_0$ and half-width-half-maximum linewidth converted from $\gamma$, respectively.
Values in the frequency and linewidth columns are resulted parameters of a center position and a half-width-half-maximum linewidth.
Numbers after the $\pm$ sign represent 95 \% confidence intervals.
The {\it Notes} column displays the algebraic relations of frequencies and energy splittings.

The outcome of algebraic relation and the center frequency can differ within the average linewidth owing to inhomogeneous broadening.
As the observed probe resonances of both the positive side and the negative side were excited by the same optical frequency for the pump laser (306.2685 THz),
they originated from the {\ybion} dopant groups of opposite resonance shifts in terms of optical inhomogeneous broadening.
Considering that both optical and spin inhomogeneous broadening are caused by a mechanical strain in the {\yso}, the probe resonances of positive and negative frequencies are seeing the reversely biased inhomogeneous broadening groups.

\newpage
\section{Optical absorption spectrum}

\begin{figure*}[h!]
\includegraphics[width=15.0cm]{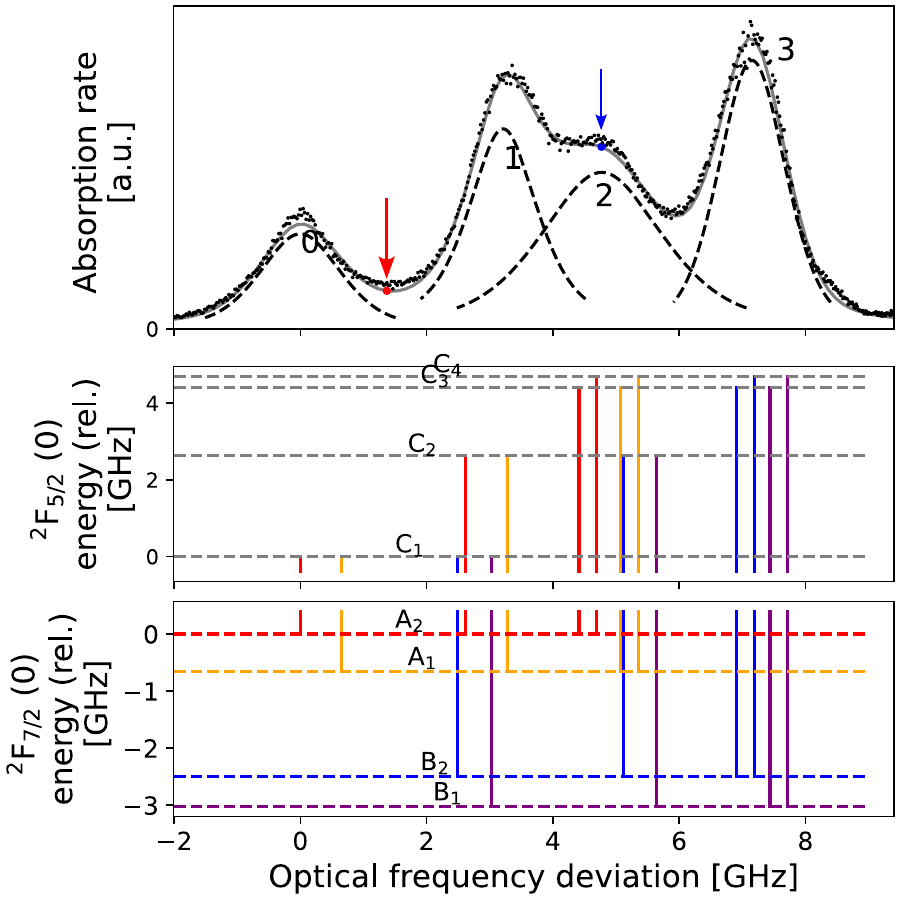}
\caption{(Color online) Optical absorption spectrum of {\ybyso} {\siteb}. The absorptions were measured for the pump laser. The horizontal axis represents optical frequency relative to 306.264 THz.
Each vertical line in the energy diagrams represents a transition between {\cflow} and {\cfhigh} excitable by the optical frequency of its location.
}\label{fig:OA}
\end{figure*}

The optical absorption spectrum of {\ybyso} {\siteb} was measured for optical frequencies from 306.264 THz (978.87 nm in wavelength) to +8 GHz of relative frequency.
The optical frequencies were obtained from a wavelength meter with a 200 MHz accuracy.
Because of optical inhomogeneous broadening between {\cflow} and {\cfhigh}, 16 transitions appeared to be 4 broad distributions ({\it 0} to {\it 3}), and their linewidths are  1.4 ({\it 0}),  1.2 ({\it 1}),  2.2 ({\it 2}), and 1.2 GHz ({\it 3}) respectively as shown as in \fref{fig:OA}.   
Comparing the optical transition energy and the absorption spectrum, $B_1$ and $B_2$ seems to have a low absorption strength for $B_2 \rightarrow C_1$ and a preference of $\{C_2, C_3, C_3\}$ for optical excitations.

The red arrow indicates an optical frequency used for the optical Raman heterodyne measurement to detect the radio-wave resonance of the $A_1 \leftrightarrow A_2$  transition, and the blue arrow used for the pump laser in the pump-probe saturation spectroscopy.

\end{document}